\documentclass[10pt,conference]{IEEEtran}
\IEEEoverridecommandlockouts
\usepackage[nobreak,space,compress]{cite}
\usepackage{amsmath,amssymb,amsfonts}
\usepackage{graphicx}
\usepackage{textcomp}
\usepackage[table]{xcolor}  
\usepackage{tcolorbox}
\def\BibTeX{{\rm B\kern-.05em{\sc i\kern-.025em b}\kern-.08em
    T\kern-.1667em\lower.7ex\hbox{E}\kern-.125emX}}
\usepackage{ulem}
\usepackage{url}
\usepackage{listings}
\usepackage{hyperref}
\usepackage{enumitem}
\usepackage{caption}
\usepackage{algorithm}
\usepackage{algpseudocode}
\usepackage{subfig}
\usepackage{balance}
\newcommand{\approach}{\textsc{MLmisFinder}}
\begin{document}


\title{MLmisFinder: A Specification and Detection Approach of Machine Learning Service Misuses}

\title{MLmisFinder: A Specification and Detection Approach of Machine Learning Service Misuses}

\author{
\IEEEauthorblockN{Hadil Ben Amor}
\IEEEauthorblockA{\textit{École de Technologie Supérieure} \\
Montréal, Canada\\
hadil.ben-amor.1@ens.etsmtl.ca}
\and
\IEEEauthorblockN{Niruthiha Selvanayagam}
\IEEEauthorblockA{\textit{École de Technologie Supérieure} \\
Montréal, Canada\\
niruthiha.selvanayagam.1@ens.etsmtl.ca}
\and
\IEEEauthorblockN{Manel Abdellatif}
\IEEEauthorblockA{\textit{École de Technologie Supérieure} \\
Montréal, Canada\\
manel.abdellatif@etsmtl.ca}
\and

\IEEEauthorblockN{\hspace{5cm}Taher A. Ghaleb}
\IEEEauthorblockA{\hspace{5cm}\textit{Trent University} \\
\hspace{5cm}Peterborough, Canada\\
\hspace{5cm}taherghaleb@trentu.ca}
\and
\IEEEauthorblockN{\hspace{0cm}Naouel Moha}
\IEEEauthorblockA{\hspace{0cm}\textit{École de Technologie Supérieure} \\
\hspace{0cm}Montréal, Canada\\
\hspace{0cm}naouel.moha@etsmtl.ca}
}

\maketitle
\begin{abstract}
Machine Learning (ML) cloud services, offered by leading providers such as Amazon, Google, and Microsoft, enable the integration of ML components into software systems without building models from scratch. 
However, the rapid adoption of ML services, coupled with the growing complexity of business requirements, has led to widespread misuses, compromising the quality, maintainability, and evolution of ML service-based systems.
Though prior research has studied patterns and antipatterns in service-based and ML-based systems separately, automatic detection of ML service misuses remains a challenge.
In this paper, we propose {\approach}, an automatic approach to detect ML service misuses in software systems, aiming to identify instances of improper use of ML services to help developers properly integrate ML components in ML service-based systems. 
We propose a metamodel that captures the data needed to detect misuses in ML service-based systems and apply a set of rule-based detection algorithms for seven misuse types.
We evaluated {\approach} on 107 software systems collected from open-source GitHub repositories and compared it with a state-of-the-art baseline.
Our results show that {\approach} effectively detects ML service misuses, achieving an average precision of 96.7\% and recall of 97\%, outperforming the state-of-the-art baseline. {\approach} also scaled efficiently to detect misuses across 817 ML service-based systems and revealed that such misuses are widespread, especially in areas such as data drift monitoring and schema validation.
\end{abstract}

\section{Introduction}
Machine Learning (ML) cloud services have quickly become central components of modern software systems, offered by leading providers such as Amazon, Google, and Microsoft. Their widespread adoption is driven by the ability to simplify ML model development, reducing the need for extensive expertise and the complexity of building models from scratch~\cite{zhang2019mark}, which has greatly facilitated their integration into software systems. Existing ML services range from extremely simple to fully customizable, supporting the development of diverse ML service-based systems~\cite{google_cloud_automl}. Their growing use for business solutions has encouraged developers of varying skill levels to adopt them to speed up system development, maintenance, and evolution~\cite{wan2021machine}. However, developers may not always follow best practices, which can lead to ML service misuses that degrade system quality and hinder maintenance and evolution~\cite{obrien202223,washizaki2022software}.
ML service misuse involves violating implicit or explicit usage constraints~\cite{benamor2025mlmisuses,wan2021machine} or applying poor practices throughout the system lifecycle. Such misuses can result in critical bugs that negatively affect the accuracy, performance, and cost-effectiveness of ML service-based systems~\cite{wan2021machine}.
For example, ML cloud providers enforce API rate limits to manage resources and prevent overload. Exceeding these limits without proper handling can lead to critical issues, such as increased latency. Similarly, failing to specify a stopping criterion during model training with an ML cloud service can increase both latency and cloud usage costs, as training may continue unnecessarily without significant performance gains~\cite{intro}. Therefore, detecting misuses of ML services is essential to improving software quality and performance.

Previous studies have explored the specification and detection of code smells and ML antipatterns from different perspectives~\cite{bogner2021characterizing,washizaki2019studying,wan2021machine,cabral2024investigating,wei2024demystifying}. However, little attention has been paid to misuses of ML services~\cite{wan2021machine,wei2024demystifying}.
Therefore, in this paper, we propose {\approach}, an automated approach to detect ML service misuses in ML service-based systems.
{\approach} is built on a novel metamodel that unifies the representation of ML service-based systems across different cloud providers to detect ML service misuses. This metamodel forms the foundation for a set of extensible static detection rules tailored to these misuses, yielding a highly accurate detection. {\approach} supports the detection of seven misuse types, most of which have not been previously addressed in the context of ML cloud services. It applies static source-code analysis without runtime data, enabling early detection during development and broad code coverage. To validate the accuracy of our findings, we analyzed 107 ML service-based systems and computed the precision and recall of {\approach}. We also compared our detection results to a state-of-the-art baseline~\cite{wan2021machine}, identifying one misuse common to both approaches.

Our results show that {\approach} effectively
detects ML service misuses, achieving an average precision of 96.7\% and recall of 97\%, outperforming the only existing state-of-the-art baseline and demonstrating its effectiveness in identifying ML service misuses, with 340 instances of misuses successfully detected across validated systems. We analyzed the prevalence of these misuses across a dataset of 817 ML service-based systems. Our findings reveal that misuses are widespread, with ``\textit{Ignoring monitoring data drift}'' and  ``\textit{Ignoring testing schema mismatch}'' occurring in 97\% and 96\% of the systems, respectively. This highlights the need for better adherence to best practices and improved developer tools for ML service integration.

The rest of this paper is organized as follows. Sections~\ref{Sec:Background} and~\ref{Sec:RW} present the background and related work, respectively. Section~\ref{Sec:approach} describes our approach. Section~\ref{Sec:studydesign} outlines our experimental setup. Empirical evaluation is
presented in Section~\ref{Sec:results}. We discuss the implications of our findings in Section~\ref{Sec:implications} and the threats to validity in Section~\ref{Sec:threats}. Section~\ref{Sec:conclusion} concludes the paper and suggests future work.

\section{Background and Related Work}
\label{Sec:prevwork}

This section provides background on the ML service misuses we aim to detect and reviews related work.

\subsection{Background}\label{Sec:Background}
In this paper, we consider seven misuses discussed in the literature~\cite{benamor2025mlmisuses,cao2022understanding,xu2022checkpointing,mallick2022matchmaker,wan2021machine}. We should note that in our prior work~\cite{benamor2025mlmisuses}, we conducted a multivocal study that resulted in a catalog of 20 ML service misuses. This catalog was derived from (a) a systematic analysis of gray and academic literature, (b) an empirical examination of ML service–based systems on GitHub, and (c) a survey with 50 ML practitioners. In this work, our selection was based on three criteria: (a) we selected misuses that cover various stages of the ML development pipeline to demonstrate that our approach is applicable across the lifecycle of ML service-based systems, (b) we prioritized misuses that are detectable through static analysis of code and service configurations, (c) we prioritized misuses having documented impact on maintainability and software quality, causing inefficiencies and technical debt. In the following, we describe the list of misuses that we detect.

\begin{itemize}[leftmargin=*]
    \item \textbf{Not using batch API for data processing}: 
    Many cloud providers offer batch processing APIs designed to improve data loading performance by handling data in batch. Despite this, developers often bypass these batch APIs, opting to load data individually or develop their own batch-processing methods. This misuse can result in out-of-memory problems, increased network traffic, and longer data loading times, which ultimately can slow down model training for example and increase operational cost~\cite{cao2022understanding,Batch}. This misuse is applicable in contexts where batch processing is required or beneficial. More specifically, in scenarios where real-time or streaming data processing is necessary, ML cloud services often provide specialized real-time APIs, and in such cases, not using a batch API would not be considered a misuse. Listing~\ref{lst:not_using_batch_api} shows an example occurrence\footnote{\url{https://github.com/ovokpus/Python-Azure-AI-REST-APIs/blob/main/text-analytics-sdk/text-analysis.py}} of this misuse and a possible fix (Listing~\ref{lst:not_using_batch_api_fix}), as the language detection batch API was not used across multiple documents.
    
\vspace{-8pt}
\begin{lstlisting}[frame=lines,caption={Not Using Batch API for Data Processing},label={lst:not_using_batch_api}]
from azure.ai.textanalytics import TextAnalyticsClient
def main():
    ...
    cog_client = TextAnalyticsClient(endpoint=cog_endpoint, credential=credential)
        # Analyze each text file in the reviews folder
        for file_name in os.listdir(reviews_folder):
            # Read the file contents
            ...
            detectedLanguage = cog_client.detect_language(documents=[text])[0]
            ...
\end{lstlisting}
\vspace{-10pt}
\begin{lstlisting}[frame=lines,caption={Not Using Batch API for Data Processing (Fix)},label={lst:not_using_batch_api_fix}]]
from azure.ai.textanalytics import TextAnalyticsClient
def main():
    ...
    # Load documents directly from files for batch calls
    documents = [open(os.path.join(reviews_folder, file_name), encoding='utf8').read() for file_name in os.listdir(reviews_folder)]
    # Batch process for language and sentiment detection
    detected_languages = cog_client.detect_language(documents=documents)
\end{lstlisting}
  
    \item \textbf{Not using training checkpoints}: Cloud providers enable resuming training from the latest checkpoint, allowing experiments to continue from their last saved state instead of restarting from scratch. This approach significantly reduces training time and computational costs, particularly for large, complex models. However, developers might overlook saving training checkpoints in cloud storage. If a model fails and no checkpoints are available, the entire training job or pipeline is forced to restart, leading to data loss since the model's progress is not retained in memory~\cite{xu2022checkpointing,rojas2020study,google2024mlbestpractices}. 

    \item \textbf{Non specification of early stopping criteria:} Many ML services offer early stopping mechanisms to prevent overfitting and minimize unnecessary computational expenses. However, developers may fail to configure these criteria, causing training to run for more epochs than necessary. This can result in excessive resource consumption, longer training durations, increased costs, and a higher risk of overfitting~\cite{microsoft2024hyperparameters,ying2019overview}.
    
     \item \textbf{Ignoring testing schema mismatch}: ML services often include features to detect discrepancies in data schemas, such as mismatches in feature or data distribution between training, testing, and production datasets, usually by triggering alerts. However, developers may overlook configuring these alerts or may choose to disable them. For instance, Amazon ML generates alerts when there is a schema inconsistency between training and evaluation data sources. Disabling these alerts can result in missed issues, such as features present in the training data but not in the evaluation data, or the inclusion of unexpected features. This oversight can lead to inaccurate model predictions and poorer performance in production environments~\cite{aws2024mlevaluationalerts,polyzotis2019data}.
   
    \item \textbf{Misinterpreting model output:} Treating the discrete, simplified outputs of ML cloud services as straightforward, despite their complex underlying semantics, can lead to critical misinterpretations and faulty application logic~\cite{wan2021machine,benamor2025mlmisuses}. Several ML cloud services return aggregated metrics (e.g., scores, magnitudes, confidence levels) from complex internal pre-trained models. A key source of misuse stems from failing to synthesize multiple output values. For example, Google's Natural Language Sentiment Analysis API returns a score (polarity) and a magnitude (intensity). Accurate classification requires both metrics~\cite{wan2021machine}. For instance,  a directional score with high magnitude implies strong sentiment (e.g., strongly positive or negative), while a score near zero or with low magnitude suggests a neutral or ambiguous result~\cite{wan2021machine}. Misinterpreting such multi-metric dependencies compromises reliability and decision quality.
          
    \item  \textbf{Improper handling of ML API limits}: 
    This occurs when API rate limits are not properly set. These limits are designed to maintain the stability and performance of the ML service. However, developers may fail to configure them correctly, leading to the sudden interruption of prediction requests once the limit is exceeded. Listing~\ref{lst:improper_handling_ml_api_limit} illustrates an example occurrence\footnote{\url{https://github.com/ramesh-15/Azure\_OpenAI\_API\_Streamlit/blob/main/azure\_openai.py}} where the allowed number of API calls could surpass the requests per second limit set by the Azure OpenAI Service. This can lead to delays or rejections until the request volume falls back within the allowed range~\cite{MicrosoftTechCommunity2024,MicrosoftOpenAIQuotasLimits}. A possible fix for this misuse is presented in Listing~\ref{lst:improper_handling_ml_api_limit_fix}.
\vspace{-12pt}
\begin{lstlisting}[frame=lines,caption={Improper Handling of ML API Limits},label={lst:improper_handling_ml_api_limit}]
import openai
def get_completion(system_message, user_message, deployment_name='deployment_name', temperature=0, max_tokens=1000) -> str:
   ...
    response = openai.ChatCompletion.create(
        engine=deployment_name,
        messages=messages,
        temperature=temperature, 
        max_tokens=max_tokens)
    return response.choices[0].message["content"]
    ...
\end{lstlisting}
\vspace{-12pt}

\begin{lstlisting}[frame=lines,caption={Improper Handling of ML API Limits (Fix)},label={lst:improper_handling_ml_api_limit_fix}]
import openai
def get_completion(system_message, user_message, deployment_name='deployment_name', temperature=0, max_tokens=1000, retries=3, delay=2) -> str:
...
  for attempt in range(retries):       
    response = openai.ChatCompletion.create(
    engine=deployment_name,messages=messages,temperature=temperature, max_tokens=max_tokens)
    return response['choices'][0]['message']['content'] 
    if attempt < retries - 1:
      print(f"Retrying in {delay} seconds...")
      time.sleep(delay)  # Wait before retrying
    else:
      print("Max retries reached, unable to get a response.")
      raise #Reraise exception if all retries fail
\end{lstlisting}    
 \item  \textbf{Ignoring monitoring for data drift}: 
    Failing to consistently monitor changes in the statistical properties or distribution of data can negatively impact model performance. Data drift occurs when the distribution of the incoming data deviates from the training data, causing a decrease in the accuracy of the model over time. Cloud providers suggest using drift and skew detection tools to track these changes and notify developers when significant shifts occur. Early detection of data drift allows for timely retraining or adjustments to the model, ensuring that it maintains optimal performance in production~\cite{bova2017intervention,mallick2022matchmaker,google2024mlgcpbestpractices2}.
\end{itemize}

\subsection{Related Work}
\label{Sec:RW}
Several approaches in the literature have been proposed to specify and detect ML (anti)patterns~\cite{masuda2018survey,bogner2021characterizing}. However, only a few of them specifically focused on the specification and detection of ML service misuses.

Wan \textit{et al.}~\cite{wan2021machine} analyzed open-source systems using Google and AWS ML APIs and identified eight ML API misuses that harm software quality. However, they focused only on cloud APIs that provide access solely to pretrained models and did not examine misuses involving data preprocessing, model training, or model deployment. In contrast to their work, our approach instead targets a different set of ML service misuses and introduces a highly automated tool to detect them.

Wei \textit{et al.}~\cite{wei2024demystifying} conducted an analysis aimed at understanding and detecting misuses of DL APIs in frameworks such as PyTorch and TensorFlow.
They developed an LLM-based API misuse detection tool, LLMAPIDet, which is designed to identify and correct these misuses.
However, their focus was primarily on DL APIs within the TensorFlow and PyTorch ecosystems, and did not extend to ML service misuses in ML service-based systems.

Shivashankar \textit{et al.}~\cite{shivashankar2025mlscent} introduced MLScent, a static analysis tool that detects ML-specific code smells across frameworks such as TensorFlow and PyTorch. They highlighted the need for ML-aware code quality tools, as traditional code smell detectors often overlook patterns that can negatively impact model performance, reproducibility, and maintainability. However, their work focuses on model-level code quality and does not address challenges specific to ML service–based systems.

Washizaki \textit{et al.}~\cite{washizaki2019studying} conducted a literature review and developer survey to collect, classify, and discuss (anti)patterns for ML-based systems. They found that developers know little about ML design patterns that could support ML system development.
Van \textit{et al.}~\cite{van2022project} defined project smells as a more holistic approach to code smells and developed \textit{mllint}, an open-source command-line tool that uses static analysis to assess the software quality of Python ML projects. However, their work does not specifically address project smells in software systems that integrate ML services.

While previous research has explored some ML (anti)patterns, a gap remains in studying ML cloud service-specific misuses. Unlike prior work focused on general ML (anti)patterns and smells, our approach specifically targets misuses in ML service-based systems. By introducing a highly automated detection method, we address key challenges in the effective integration and use of ML services.

\section{Approach}
\label{Sec:approach}

\begin{figure}
  \centering
  \includegraphics[width=0.94\linewidth]{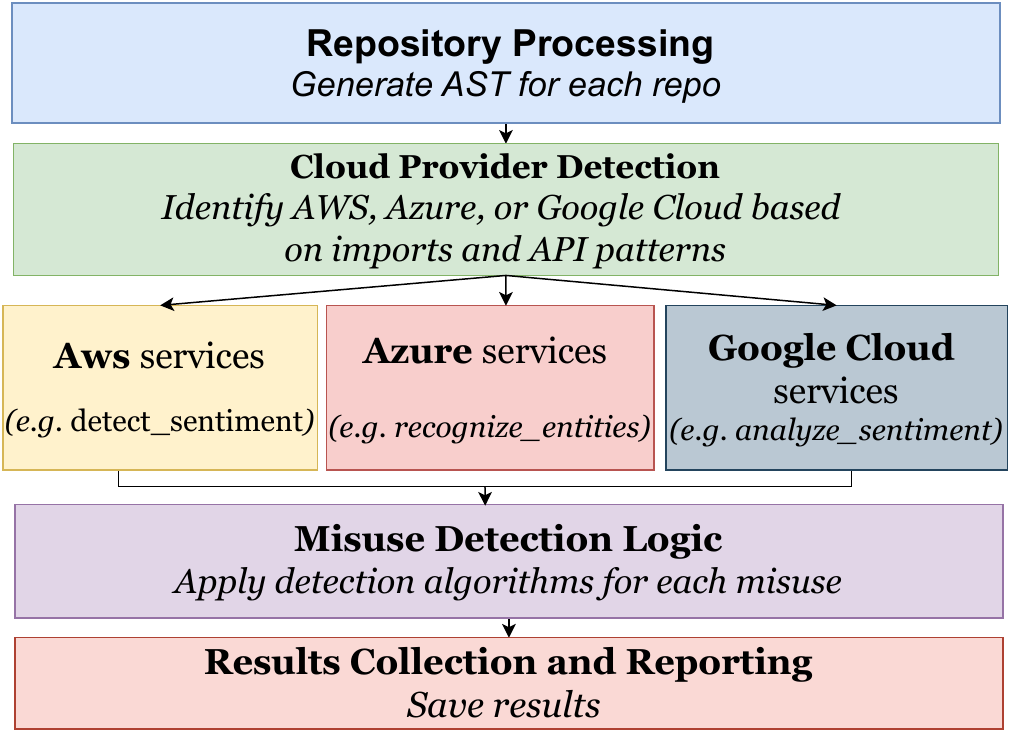}
  \caption{Overview of {\approach}}
  \label{fig:approach}
  \vspace{-10pt}
\end{figure}

In this section, we present {\approach}, a fully automated approach for detecting the ML service misuses described in Section~\ref{Sec:Background}. We should note that our detection approach supports three major ML cloud providers: AWS, Azure, and Google. As shown in Figure~\ref{fig:approach}, {\approach} takes the GitHub repository of an ML service-based system as input, clones it, and parses its source code to extract the data needed for misuse detection. This data is structured using a dedicated metamodel, which serves as the basis for the detection process. To populate the metamodel, {\approach} adopts parsers to generate an Abstract Syntax Tree (AST), analyzes the source code, and derives the required metamodel constituents. The metamodel is then instantiated to build a concrete project model for analysis.
Once the metamodel is instantiated, {\approach} applies a set of detection rules (referred to hereafter as detection algorithms) to the model's constituents to identify and count occurrences of each specified ML service misuse.
Combining static analysis, metamodel instantiation, and dedicated detection algorithms, {\approach} offers the first fully automated approach for detecting such ML service misuses in software systems. In the following, we describe our metamodel and the detection algorithms for the seven ML service misuses that {\approach} detects.

\subsection{Metamodel Definition}

\begin{figure}[t]
  \centering
  \includegraphics[width=1\linewidth]{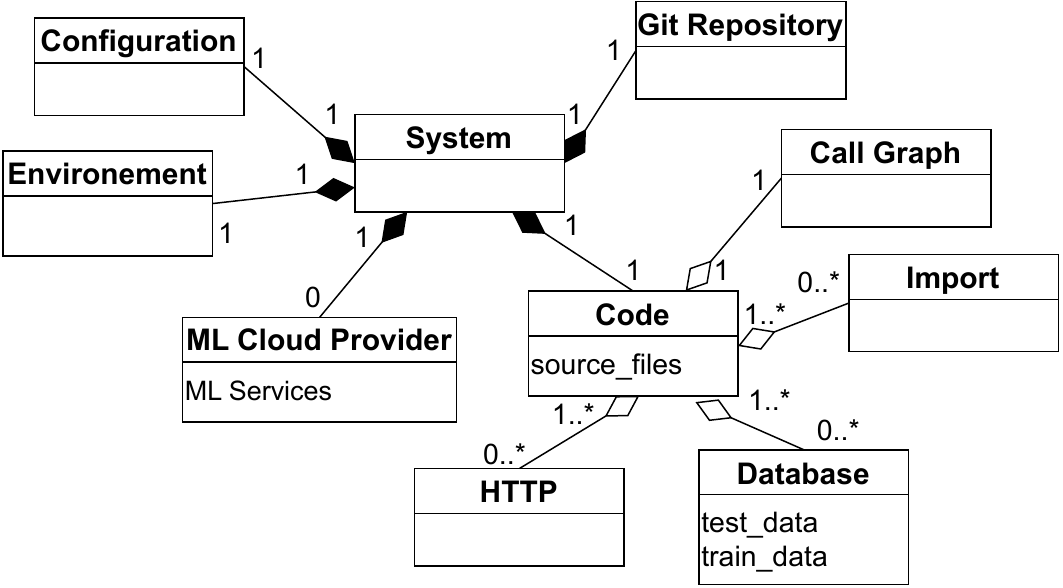}
  \caption{Metamodel Constituents in {\approach}}
  \label{fig:metamodel}
  \vspace{-10pt}
\end{figure}

We developed a metamodel to describe the data required by our detection algorithms, ensuring adaptability across diverse environments and use cases. Designed for extensibility, it separates key concepts such as data sources, processing steps, ML services, and configuration parameters, allowing new misuse types to be captured by extending or recomposing existing elements. The metamodel defines the core detection components and their relationships, offering a structured representation of the system’s environment, configuration, codebase, cloud services, and data sources. Figure~\ref{fig:metamodel} illustrates the constituents of our metamodel and their associations.

\subsection{Metamodel Constituents}

\begin{itemize}
    \item \textbf{\texttt{System}}: This is the root of the metamodel. It represents the ML service-based application and its operational context. It interacts with key external and internal constituents such as the \texttt{Environment}, \texttt{Configuration}, \texttt{Git Repository}, \texttt{ML Cloud Provider}, and \texttt{Code}.

    \item \textbf{\texttt{Environment}}: Represents the overall context in which the \texttt{System} operates and manages data about the commonly used environment variables that are dynamically injected into a system.
    
    \item \textbf{\texttt{Configuration}}: Stores data gathered from the configuration files of the ML services and defines how the \texttt{System} and its services are set up, such as enabled/disabled features.
    
    \item \textbf{\texttt{ML Cloud Provider}}: Represents external cloud platforms that provide ML services, such as AWS, Azure, or Google Cloud, which the \texttt{System} uses. It offers various \texttt{ML Services}, including model training, inference, and data processing, making them critical points for understanding external interactions and potential misuse risks.
    
    \item \textbf{\texttt{Git Repository}}: The version-controlled storage that holds the \texttt{System}'s \texttt{Code} (\texttt{source\_files}) that is analyzed to detect potential misuses.
    
    \item \textbf{\texttt{Code}}: Consists of the \texttt{System}'s source files and forms the backbone of misuse detection. It provides insight into how the system operates and interacts with external components. Key elements include:
    
    \begin{itemize}
        \item \texttt{Import}: A list of libraries and dependencies used in the source code, which can affect system behavior and security.
        
        \item \texttt{HTTP}: Records external HTTP requests made by the \texttt{System}, often for ML service calls or other network interactions, useful for tracking data flow.
        
        \item \texttt{Database}: Represents the storage layer, including \texttt{train\_data} and \texttt{test\_data}, which are critical for model evaluation and potential targets for misuse.
        
        \item \texttt{Call Graph}: A representation of function calls within the \texttt{Code}, derived from an AST, helping to analyze system interactions and detect risky pathways.
    \end{itemize}
    
\end{itemize}

\subsection{Metamodel Instantiation}
\vspace{-3pt}
{\approach}'s metamodel is instantiated by analyzing various aspects of the ML service-based system, as follows.

\begin{enumerate}
    \item \textbf{\textit{Code Extraction}}: The \texttt{System}'s source code is parsed using Python’s \textit{ast} module to extract function calls, control flow structures, and import statements.
    
    \item \textbf{\textit{Call Graph Construction}}: An AST is used to generate a \texttt{Call Graph}, capturing function relationships and interactions.

     \item \textbf{\textit{ML Cloud Provider Identification}}: The \texttt{System} is checked for predefined invocation patterns associated with cloud-based ML platforms such as AWS, Azure, and Google Cloud, including \texttt{\textit{boto3}}, \texttt{\textit{azureml}}, and \texttt{\textit{vertexai}}, respectively. This helps establish a contextual foundation that guides the detection of misuses.
    
    \item \textbf{\textit{Database Analysis}}: Training and testing datasets are identified by detecting function calls such as \texttt{train\_test\_split}, \texttt{train}, \texttt{fit}, \texttt{predict}, and \texttt{evaluate}.
    
    \item \textbf{\textit{HTTP Request Detection}}: External API calls to ML services are identified and analyzed for potential misuse.
    
    \item \textbf{\textit{Configuration and Environment Analysis}}: Configuration files and environment variables are examined to identify relevant framework settings and execution parameters.
\end{enumerate}

\subsection{Detection Algorithms}
\vspace{-2pt}
For each ML service misuse, we defined a detection algorithm to detect its occurrences in a given ML service-based system. Once a misuse is detected, {\approach} generates a report containing the repository name, misuse type, number of occurrences, and the specific line of code where the misuse occurred, if applicable (e.g.,  the case of ``\textit{Not Using Batch API for Data Processing}''). In the following, we provide a detailed description of each detection algorithm. 

\vspace{4pt}
\subsubsection{Not Using Batch API for Data Processing}
This misuse occurs when batch-processing APIs are available but inefficiently used within loops instead of processing multiple inputs in a single request. The detection algorithm verifies the \texttt{ML Cloud Provider} used, examines API calls in \texttt{Code}, and checks the \texttt{Call Graph} to determine if batch APIs are invoked inside loops, as illustrated in Algorithm~1.

\linespread{0.9}
\begin{algorithm}[ht]
\small
\caption{Not Using Batch API for Data Processing}
\begin{algorithmic}[1]
\State \textbf{Start:} Analyze ML API calls in \texttt{Code}
\State Compare API calls to known batch-processing APIs
\If {API supports batch processing}
    \State Check if API is used inside loop(s)
    \If {Batch API is used inefficiently}
        \State \textit{Flag misuse}
    \Else
        \State \textit{No misuse detected}
    \EndIf
\Else
    \State \textit{No misuse detected}
\EndIf
\end{algorithmic}
\end{algorithm}

\subsubsection{Not Using Training Checkpoints}
Checkpointing helps prevent loss of progress during training interruptions. The detection algorithm examines the \texttt{Import} constituent for \texttt{ML Cloud Provider} to determine whether a training component is present in the system. If identified, the algorithm proceeds to further examine the \texttt{Call Graph} to track checkpoint-saving and restoration functions, and ensures checkpoints are correctly implemented in \texttt{Code}, as demonstrated in Algorithm~2.

\vspace{-5pt}
\linespread{0.9}
\begin{algorithm}[ht]
\small
\caption{Not Using Training Checkpoints}
\begin{algorithmic}[1]
\State \textbf{Start:} Analyze ML Cloud Provider SDK
\If {No SDK detected}
\State \textit{No misuse detected}
\Else
\State Analyze checkpoint saving/restoration calls
\If {No checkpointing functions are found}
\State \textit{Flag misuse}
\Else
\If {Checkpoints are saved but never restored}
\State \textit{Flag misuse}
\Else
\State \textit{No misuse detected}
\EndIf
\EndIf
\EndIf
\end{algorithmic}
\end{algorithm}

\vspace{4pt}
\subsubsection{Non Specification of Early Stopping Criteria}
Early stopping is crucial to prevent overfitting. The detection algorithm begins by checking whether any training component exists in the \texttt{Code}. If present, it subsequently verifies whether early stopping-related libraries are imported in \texttt{Import}, analyzes function calls in \texttt{Call Graph}, and ensures that proper stopping criteria are set in \texttt{Code}, as illustrated in Algorithm~3.
\vspace{-10pt}
\linespread{0.9}
\begin{algorithm}
\small
\caption{Non Specification of Early Stopping Criteria}
\begin{algorithmic}[1]
\State \textbf{Start:} Analyze ML Cloud Provider SDK
\If {No SDK detected}
    \State \textit{No misuse detected}
\Else
    \State Analyze early stopping libraries
    \If {No early stopping library is imported}
        \State \textit{Flag misuse}
    \Else
        \State Analyze early stopping function usage
        \If {Early stopping is not properly configured}
            \State \textit{Flag misuse}
        \Else
            \State \textit{No misuse detected}
        \EndIf
    \EndIf
\EndIf
\end{algorithmic}
\end{algorithm}

\vspace{-2pt}
\subsubsection{Ignoring Testing Schema Mismatch}
ML pipelines should validate schema consistency between training and testing datasets. The detection algorithm inspects the \texttt{Import} constituent for validation libraries, checks whether schema validation functions are used in \texttt{Code}, and examines the \texttt{Database} to determine if train and test datasets are explicitly compared for schema consistency, as depicted in Algorithm~4.
\vspace{-5pt}
\linespread{0.9}
\begin{algorithm}[H]
\small
\caption{Ignoring Testing Schema Mismatch}
\begin{algorithmic}[1]
\State \textbf{Start:} Analyze \texttt{Import} constituent for schema validation libraries
\If {Validation libraries are missing}
\State \textit{Flag misuse}
\Else
\State Analyze \texttt{Code} for validation function usage
\If {Validation functions are not used}
\State \textit{Flag misuse}
\Else
\State Check if train and test data schemas are compared
\If {No schema comparison is performed}
\State \textit{Flag misuse}
\Else
\State \textit{No misuse detected}
\EndIf
\EndIf
\EndIf
\end{algorithmic}
\end{algorithm}

\subsubsection{Misinterpreting Output} analyzes the \texttt{Code} and its \texttt{Call Graph} to see how the model’s outputs are used. It identifies all output-related API methods or properties, determines which ones are actually used in the application, and checks whether the code relies on only a subset of the outputs needed for correct interpretation. If the code uses a single output (e.g., \texttt{.score}) while ignoring other required outputs (e.g., \texttt{.magnitude}), it flags a potential misinterpretation of the model output, as illustrated in Algorithm~5.

\linespread{0.9}
\begin{algorithm}
\small
\caption{Misinterpreting Output}
\begin{algorithmic}[1]
\State \textbf{Start:} Analyze \texttt{Code} and \texttt{Call Graph}
\If {No output methods are called}
    \State \textit{No misuse detected}
\Else
    \If {The code uses only a subset of the model’s output values}
        \State \textit{Flag misuse}
    \Else
        \State \textit{No misuse detected}
    \EndIf
\EndIf
\end{algorithmic}
\end{algorithm}
\subsubsection{Improper Handling of ML API Limits}
ML services impose API limits that should be handled to avoid failures. The detection algorithm examines \texttt{Import} for monitoring libraries, verifies their usage in \texttt{Code}, inspects the \texttt{Call Graph} for relevant monitoring functions, and checks \texttt{HTTP} requests for rate limit monitoring by analyzing request headers and parameters, as outlined in Algorithm~6.

\linespread{0.9}
\begin{algorithm}[ht]
\small
\caption{Improper Handling of ML API Limits}
\begin{algorithmic}[1]
\State \textbf{Start:} Analyze monitoring libraries
\If {No monitoring library is imported}
\State \textit{Flag misuse}
\Else
\State Analyze monitoring function usage
\If {Monitoring functions are not used}
\State \textit{Flag misuse}
\Else
\State Analyze \texttt{HTTP} requests for rate limit monitoring
\If {No rate limit checks are performed}
\State \textit{Flag misuse}
\Else
\State \textit{No misuse detected}
\EndIf
\EndIf
\EndIf
\end{algorithmic}
\end{algorithm}

\subsubsection{Ignoring Monitoring for Data Drift}
ML models should be monitored for data drift to ensure continued accuracy. The detection algorithm analyzes the \texttt{Import} constituent for data drift monitoring libraries, verifies their application in \texttt{Code}, and ensures appropriate data drift monitoring functions are instantiated, as shown in Algorithm~7.

\linespread{0.9}
\begin{algorithm}[ht]
\small
\caption{Ignoring Monitoring for Data Drift}
\begin{algorithmic}[1]
\State \textbf{Start:} Analyze drift monitoring libraries
\If {No monitoring libraries are imported}
\State \textit{Flag misuse}
\Else
\State Analyze \texttt{Code} for drift monitoring function usage
\If {Drift monitoring functions are not used}
\State \textit{Flag misuse}
\Else
\State \textit{No misuse detected}
\EndIf
\EndIf
\end{algorithmic}
\end{algorithm}


\section{Experimental Setup}
\label{Sec:studydesign}
We now present the study design used to validate our approach. We applied {\approach} to a set of ML service-based systems and compared the detected occurrences with a ground truth (i.e., the manually identified misuses in the systems). Below, we describe the baseline approaches, evaluation dataset, ground truth construction, and research questions.

\vspace{-5pt}
\subsection{Baseline} \label{Baseline}
\vspace{-3pt}
Since our main focus is on detecting ML service misuses, we selected baseline approaches for comparison based on three criteria: (1) it must detect at least one ML service misuse {\approach} detects, to align with our objective; (2) it must be peer-reviewed with open-source implementation for replicability and enable a fair comparison with {\approach}; (2) it must support ML cloud services and at least one of the major cloud providers in our study.
Based on the first two criteria, we found Wan \textit{et al.}'s approach~\cite{wan2021machine} and MLScent~\cite{shivashankar2025mlscent} to be relevant for our misuse categories.
Wan \textit{et al.}~\cite{wan2021machine} detect misuses in the usage of ML APIs from Google and AWS, thus meeting our third selection criterion. The only misuse we share in common with their work is ``\textit{Misinterpreting Output}'', which aligns directly with our evaluation goals.
MLScent is a static analysis tool for detecting ML antipatterns using frameworks such as TensorFlow and PyTorch. It supports the ``\textit{Non-specification of Early Stopping Criteria}'' misuse, which is the only misuse shared with our work. However, we excluded it since it only detects this misuse at the framework level, not at the ML cloud service level, failing to satisfy the third criterion.
Therefore, we retained only Wan \textit{et al.}'s approach~\cite{wan2021machine} to compare with {\approach}.

\subsection{Dataset} \label{sec:data}
\begin{figure*}[ht]
  \vspace{-7pt}
  \centering
  \subfloat[Lines of Code]{
    \includegraphics[width=0.30\textwidth]{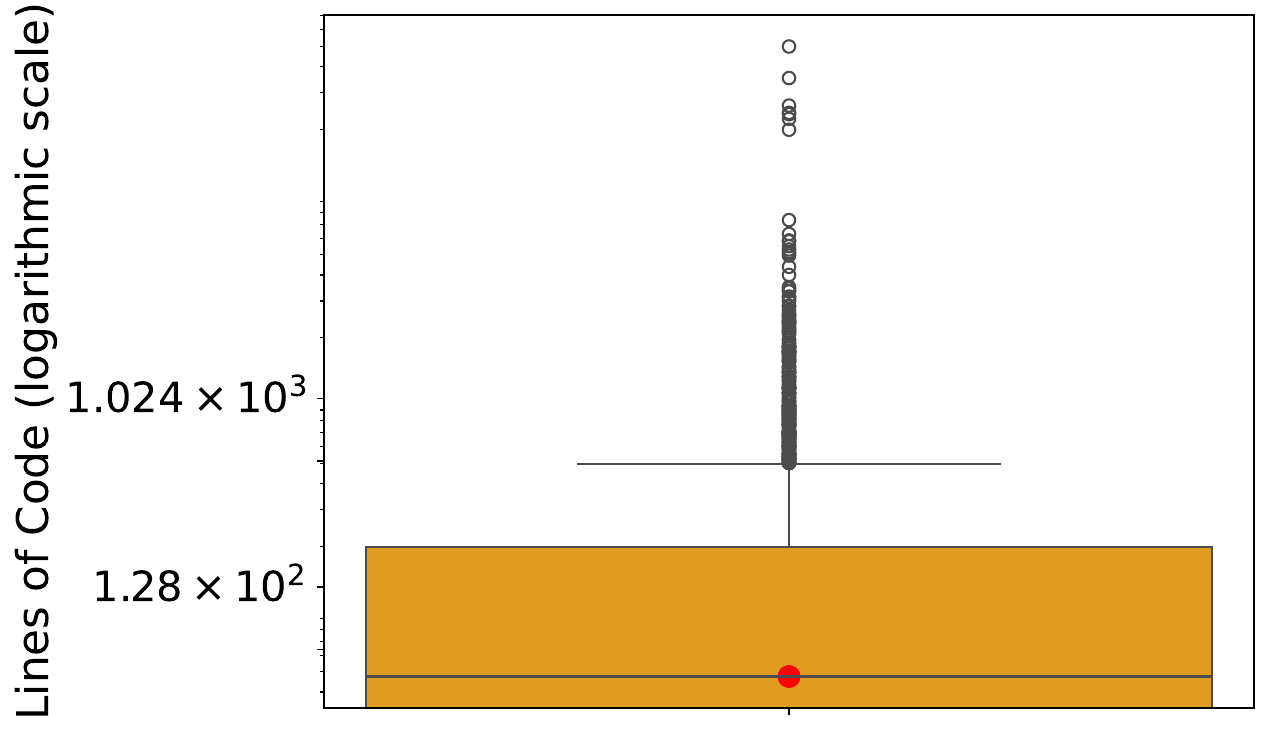}
    \label{fig:lines_of_code}
  }
  \subfloat[Number of Files ]{
    \includegraphics[width=0.26\textwidth]{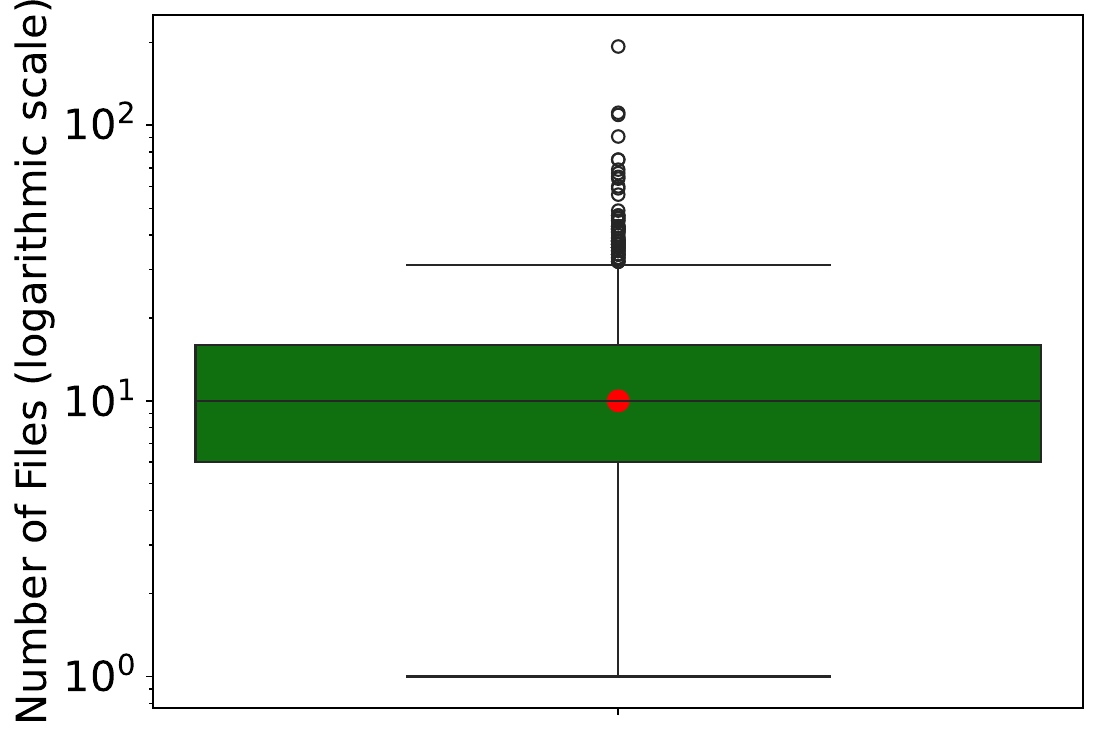}
    \label{fig:number_of_files}
  }
  \subfloat[Stars and Forks]{
    \includegraphics[width=0.23\textwidth]{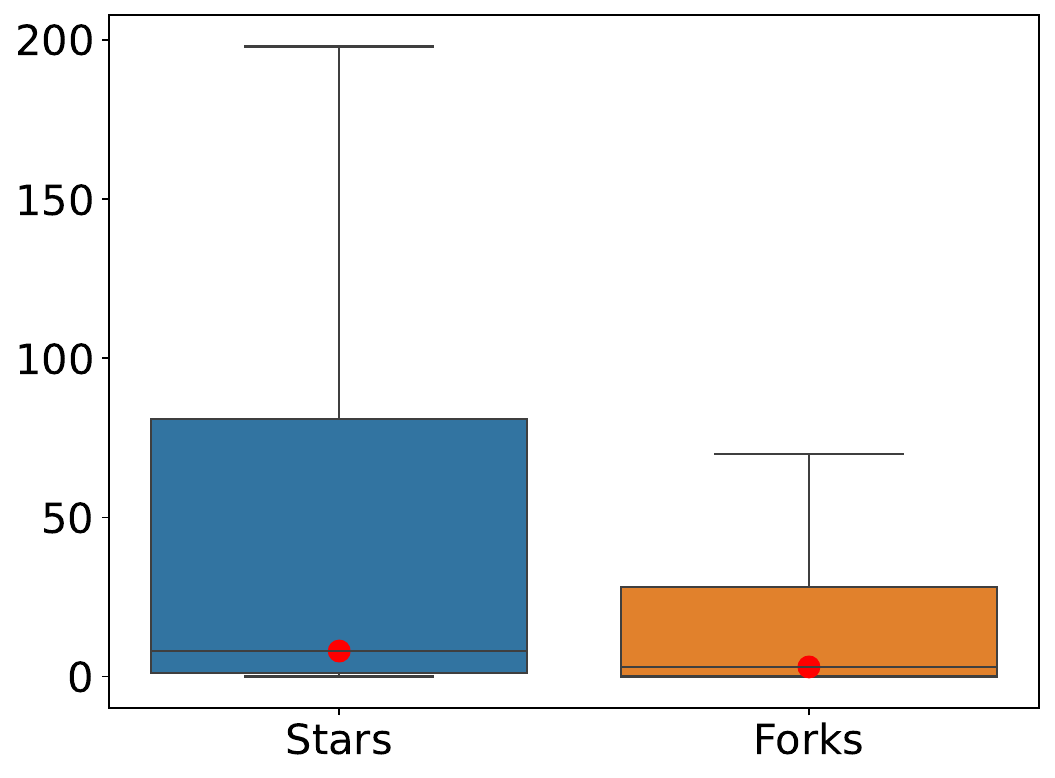}
    \label{fig:stars_forks}
  }
  \vspace{-3pt}
  \caption{Statistical Distribution of GitHub Repository Metrics}
  \label{fig:dist}
  \vspace{-9pt}
\end{figure*}

\vspace{-3pt}
We collected open-source ML service-based projects from GitHub over four months, with no restriction on repository creation dates. We used the GitHub API to automate the retrieval process, allowing structured and efficient access to repository metadata. We used service-specific keywords from cloud provider documentation, such as ``\textit{ML cloud}'', ``\textit{Azure cognitive service}'', ``\textit{API}'', and ``\textit{Azure AI}'' to identify Azure ML repositories. In addition, we used GitHub’s search functionality to detect the presence of these keywords within Python files. Initial data collection yielded 817 GitHub repositories primarily written in Python and related to ML services. We then applied an automated filtration process by analyzing project descriptions and source code to assess relevance. This included checking the presence of Jupyter notebooks and the inclusion of ML-related Python modules. Such criteria refined the dataset to projects actively using ML services, ensuring the quality and focus of our subsequent analysis. All the collected project metadata is available in our replication package~\cite{replication}.
Figure~\ref{fig:dist} shows statistics on the collected systems, which vary in size, star count, and number of forks.

\subsection{Ground Truth}
To evaluate {\approach}'s detection accuracy, we manually constructed a ground truth dataset of ML service misuses. Reviewing all 817 projects, corresponding to 5,719 individual checks per evaluator (seven misuse types per project), would have required significant time and effort. To address this, we employed statistical random sampling to select a representative subset of projects from our collected dataset of open-source ML service-based systems on GitHub (Section~\ref{sec:data}). With a $\pm 10$ margin of error and a 95\% confidence level, we determined a required sample size of 87, which ensures reliable accuracy assessment while remaining feasible for manual analysis. Nevertheless, we considered more projects and randomly analyzed a total of 107 samples to further enhance the reliability and depth of our findings. Three evaluators experienced in ML services and software engineering, manually and independently analyzed the 107 randomly selected projects, to identify instances of the seven misuses.
We then calculated Cohen's Kappa coefficient, which reached 84.7\%, indicating strong agreement in the manual detection process. This evaluation identified 340 occurrences of the seven ML service misuses, making our dataset the most comprehensive ground truth collection for ML service misuses, to the best of our knowledge. To facilitate further research, we have publicly released this dataset as part of our replication package~\cite{replication}.

\section{Empirical Evaluation}
\label{Sec:results}

Our empirical evaluation aims to address three specific research questions.

\subsection{\textbf{RQ1. How effective is {\approach} in detecting ML service misuses?}}
\noindent\textbf{Motivation.} 
We aim to evaluate the accuracy of {\approach} in identifying the seven misuses through static analysis of ML service-based systems. Specifically, we aim to measure its ability to detect various misuse types across GitHub repositories and compare its detection results with the state-of-the-art baseline.

\vspace{2pt}
\noindent\textbf{Results.} 
To evaluate the effectiveness of {\approach} in detecting ML service misuses, we applied it to a statistical sample of 107 ML service-based systems to measure its accuracy in identifying such misuses. We compared the detection results with the manually curated ground truth to evaluate precision, recall, and F1 score.

Our empirical evaluation revealed that {\approach} consistently achieved high precision and recall, demonstrating its robustness in identifying ML service misuses. A detailed breakdown of detection results across the different misuses is presented in Table~\ref{tab:res_time}. Specifically, {\approach} achieved precision ranging from 80\% to 100\%, with an average of 96.7\%, and values ranging from 76.2\% to 100\%, with an average of 97\%.
These results indicate that {\approach} correctly detects a large number of ML service misuses while maintaining a low false positive rate.
We observe that ``\textit{Ignoring monitoring for data drift}'' was detected with perfect precision and recall, meaning that {\approach} identified all true instances of this misuse without any false positives, highlighting the reliability of our detection algorithm. Certain misuses, such as ``\textit{Not using training checkpoints}'' and ``\textit{Non specification of early stopping criteria}'' were detected with relatively high precision (80\% and 81\%, respectively) and perfect recall, meaning that {\approach} detected every actual misuse of these types while occasionally misclassifying a few non-misuse cases, underscoring the robustness and reliability of our detection algorithms for these particular types of misuses. ``\textit{Not using batch API for data processing}'' exhibited a slightly lower recall of 90\% compared to other misuses, mainly attributed to a limitation in detecting deprecated or unknown ML APIs from cloud providers for batch processing, which are not supported by {\approach} as their documentation has been retired, potentially raising another type of warning for ML engineers. {\approach} also incorrectly identified two occurrences of the same misuse because of the problem of linked functions, where an ML API is called inside a function that is iterated over in a loop. {\approach} may fail to recognize that the ML API is repeatedly invoked in individual iterations instead of processing in an optimized batch mode, especially when the linked functions are complex.

\vspace{2pt}
\noindent\textbf{Comparison with \textit{Output Misinterpretation Checker}~\cite{wan2021machine}.} To benchmark {\approach}, we ran the \textit{Output Misinterpretation Checker} from Wan \textit{et al.}~\cite{wan2021machine} on a subset of our dataset.
While {\approach} supports services from three major cloud providers, Wan \textit{et al.}’s tool supports only Amazon and Google APIs. For a fair comparison, we ran their tool on the corresponding subset of 74 repositories (out of 107). It successfully processed 68, while six failed with \textit{error \texttt{-2}} due to files exceeding 1,000 lines, a known limitation noted in the original study. On valid samples, the baseline achieved 17.3\% precision, 56.2\% recall, and a 26.5 F1-score. While recall was moderate, the low precision and F1-score highlight the limitations of its rigid detection rules and limited practical reliability. For example, the tool relies on hard-coded heuristics tightly coupled with specific Google ML APIs, restricting its applicability and generalizability to other cloud providers.

In contrast, {\approach} uses modular, extensible rules derived from its underlying metamodel of ML service usage and grounded in common misuse patterns across APIs. It performs reliably without file size limits, enabling broader applicability and more accurate misuse detection. {\approach} uses regular expressions that span different cloud providers, with patterns broad enough to capture API variations (e.g., \textit{.score}, \textit{.Sentiment}, \textit{.sentiment}). This generality can lead to false negatives when relevant checks are spread across multiple lines. These challenges reveal our approach’s limitations and show that, although {\approach} mitigates some issues via a structured, extensible metamodel, fully addressing them requires deeper semantic understanding of code behavior.

\begin{table*}[ht]
\centering
\small
\caption{Detection Performance Results and Average Execution Time: \approach\ vs. Wan et al.~\cite{wan2021machine}}
\captionsetup{skip=4pt}
\setlength{\tabcolsep}{4pt}
\renewcommand{\arraystretch}{1.2}
    \begin{tabular}{|l|cccc|cccc|}
    \hline
    \rowcolor{gray!30}
    \textbf{Misuse Type} & \multicolumn{4}{c|}{\textbf{\approach}} & \multicolumn{4}{c|}{\textbf{Wan \textit{et al.}~\cite{wan2021machine}}} \\
    \rowcolor{gray!15}
    & Precision & Recall & F1 & Avg Time (s) & Precision & Recall & F1 & Avg Time (s) \\
    \hline
    Misinterpreting output & 100\% & 76.2\% & 86.5\% & 8.17 & 17.3\% & 56.2\% & 26.5\% & 70 \\
    \hline
    Not using Batch API for Data Processing & 90\% & 90\% & 90\% & 34.73 & \multicolumn{4}{c|}{Not Supported} \\
    Not using training checkpoints & 80\% & 100\% & 88.9\% & 1.06 & \multicolumn{4}{c|}{Not Supported} \\
    Non-specification of early stopping criteria & 81\% & 100\% & 89.5\% & 1.07 & \multicolumn{4}{c|}{Not Supported} \\
    Ignoring testing schema mismatch & 99\% & 100\% & 99.5\% & 1.21 & \multicolumn{4}{c|}{Not Supported} \\
    Improper handling of ML API limits & 100\% & 92.6\% & 96.2\% & 1.20 & \multicolumn{4}{c|}{Not Supported} \\
    Ignoring monitoring for data drift & 100\% & 100\% & 100\% & 1.18 & \multicolumn{4}{c|}{Not Supported} \\
    \hline\hline
    \textbf{Overall} & \textbf{96.7\%} & \textbf{97\%} & \textbf{96.8\%} & \textbf{50.05} & \textbf{12.2\%} & \textbf{56.2\%} & \textbf{26.5\%} & \textbf{70} \\
    \hline
    \end{tabular}
\label{tab:res_time}
\vspace{-2pt}
\end{table*}

\begin{tcolorbox}[top=2pt, bottom=2pt, 
  left=2.2pt, right=2.2pt, boxsep=0pt, ]\small
  \textbf{Answer to RQ1:} {\approach} demonstrated a high effectiveness in detecting ML service misuses, achieving an average precision of 96.7\% and a recall of 97\%. This confirms its reliability as a robust static analysis approach for identifying ML service misuses. It also significantly outperformed the existing baseline in detecting   ``\textit{Output Misinterpretation}'', the only misuse supported by both approaches. 
\end{tcolorbox}

\vspace{-2pt}
\subsection{\textbf{RQ2. How efficient is {\approach} in detecting ML service misuses?}}
\noindent\textbf{Motivation.} Our objective is to evaluate the efficiency of {\approach} in terms of execution time when detecting misuse of ML services between projects of different sizes and to assess the scalability of our approach. Specifically, we aim to measure how {\approach}'s performance is affected by factors, such as the number of source files and lines of code within a repository.

\begin{figure}[ht]
  \vspace{-10pt}
  \centering

  \subfloat[Execution Time vs. Lines of Code]{
    \includegraphics[width=0.9\linewidth, height=3cm]{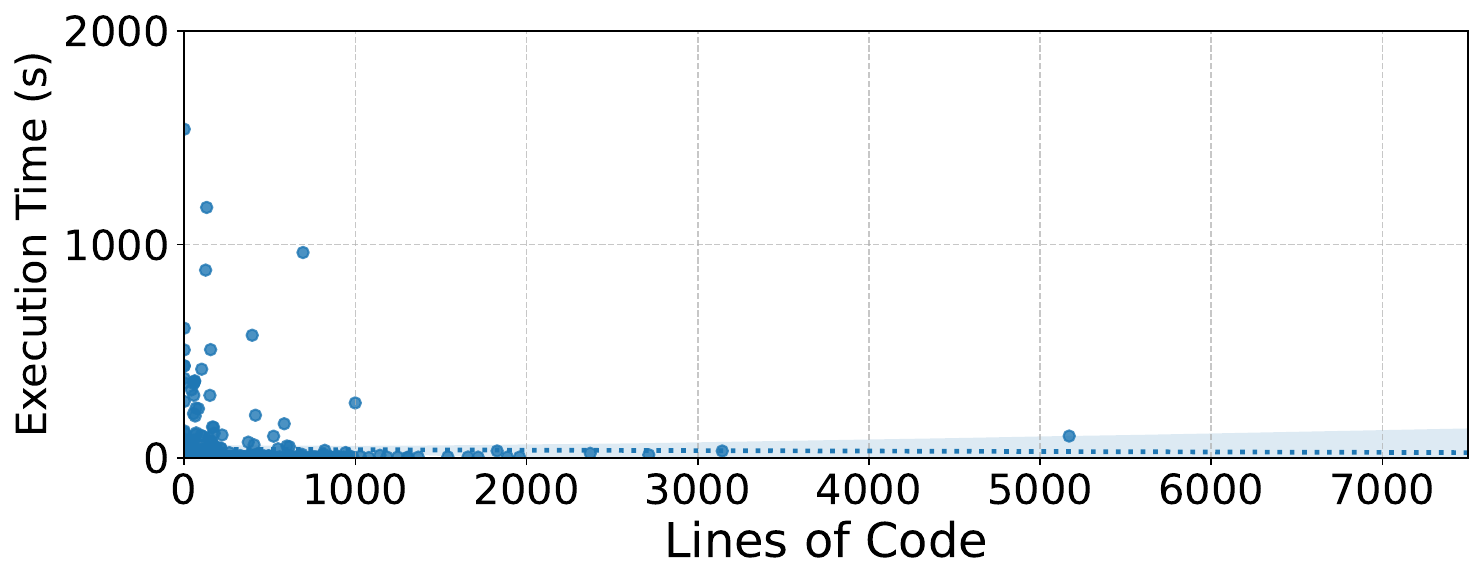}
    \label{fig:loc}
  }
  \vspace{-12pt}
  \subfloat[Execution Time vs. Number of Files]{
    \includegraphics[width=0.9\linewidth,height=3cm]{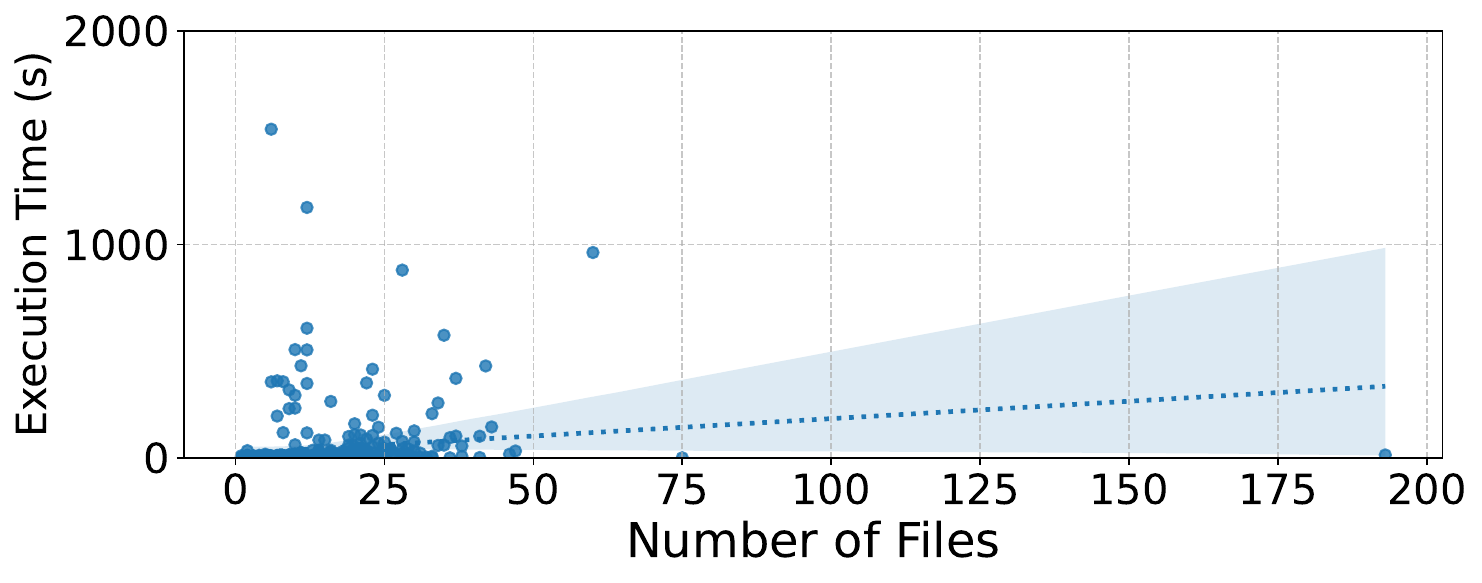}
    \label{fig:numfilesTime}
  }
  \caption{Execution Time for Different Projects Characteristics}
  \vspace{-10pt}
\end{figure}

\begin{figure}[ht]
  \vspace{-5pt}
  \centering
  \includegraphics[width=1.1\linewidth]{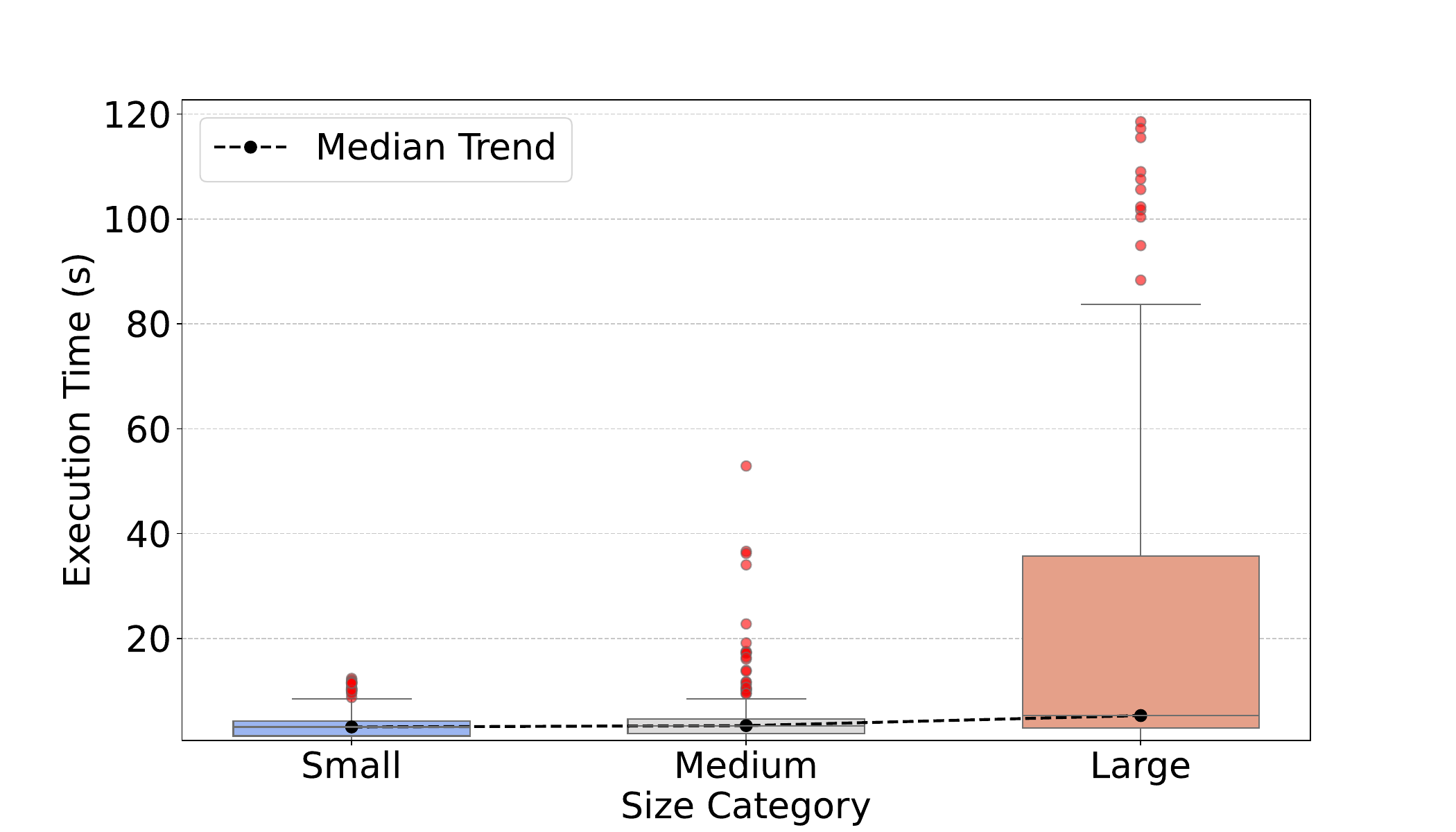}
  \caption{Execution Time Variability Across Projects Size}
  \label{fig:execution_time}
\end{figure}

\vspace{3pt}
\noindent\textbf{Results.} To assess the efficiency of {\approach} in detecting ML service misuses, we measured its execution time across all 817 projects in our dataset. We also evaluated the execution time for each individual misuse detection algorithm, excluding the time required for metamodel instantiation. The results (presented in Table \ref{tab:res_time}) confirm that {\approach} is computationally efficient and scalable, making it practical for real-world use, where timely feedback during development is important. 
The average execution time of {\approach} across the validation projects (up to 19,879 LOCs) is about $50$ seconds. As shown in Figures~\ref{fig:loc} and~\ref{fig:numfilesTime}, execution time scales linearly with LOCs and the number of files, ensuring efficient detection even in large-scale systems. Specifically, while larger projects naturally require more processing time, the increase remains moderate and does not exhibit exponential growth, allowing {\approach} to be used effectively across projects of varying complexity and size. 
We also analyzed variability in execution time across projects of different sizes. As shown in Figure~\ref{fig:execution_time}, small ($\leq 4,045$ LOCs) and medium-sized ($4,046–5,169$ LOCs) projects show minimal fluctuations, whereas larger systems ($\geq 5,170$ LOCs) exhibit greater variability, with some projects requiring longer analysis. In practice, this means that developers can rely on {\approach} for timely detection of ML service misuses, even in large or complex systems, without significant workflow delays. Table \ref{tab:res_time} also shows that most misuse detection algorithms execute in approximately one second, highlighting the efficiency of {\approach}. For ``\textit{Misinterpreting Output}'', {\approach} takes around 8 seconds, roughly nine times faster than the \textit{Output Misinterpretation Checker}~\cite{wan2021machine}, further demonstrating its superior performance. The detection of ``\textit{Not Using Batch API for Data Processing}'' is more time-consuming, averaging 35 seconds, primarily due to call graph analysis and loop structure evaluation, which require identifying specific ML service invocation patterns and assessing inefficient use of batch processing. Nevertheless, the computational overhead remains manageable, and the longer execution time is justified by the complexity of the misuse being detected.

\begin{tcolorbox}[top=2pt, bottom=2pt, 
  left=2.2pt, right=2.2pt, boxsep=0pt, ]\small
  \textbf{Answer to RQ2:} 
{\approach} is efficient and scalable, with execution time scaling linearly with project size. Most algorithms run in about one second, except for ``\textit{Not Using Batch API for Data Processing}''. Despite some variation in larger projects, execution time remains low, supporting real-world applicability.
\end{tcolorbox}

\vspace{-4pt}
\subsection{\textbf{RQ3. How prevalent are ML service misuses in ML service-based systems?}}

\noindent\textbf{Motivation.} We aim to analyze the distribution of ML service misuses across different ML service-based systems. Specifically, we aim to understand how frequently different types of misuses occur and check whether certain misuses are more prevalent in real-world projects.

\vspace{3pt}
\noindent\textbf{Results.} Running {\approach} across the 817 repositories revealed a widespread but variable presence of ML service misuses, with some occurring at notably high rates. ``\textit{Ignoring monitoring for data drift}'' was the most frequently detected misuse, appearing in 98\% of repositories (803 occurrences), followed closely by ``\textit{Ignoring testing schema mismatch}'', found in 97\% of projects (789 occurrences). These results suggest that many ML service-based systems do not explicitly implement mechanisms to track data distribution shifts or ensure schema consistency across different stages of the ML development pipeline.  
However, possible limitations in detection scope should be considered. The absence of data drift monitoring in the source code does not necessarily imply it is completely ignored; it could be implemented in external monitoring services, logging mechanisms, or the infrastructure layer, which static analysis may not capture. Similarly, schema validation may not always be relevant, particularly for projects using simple or standard datasets, where developers may not find explicit schema checks necessary, potentially explaining the high prevalence of this detected misuse.

Other misuses, while less common, still occur at notable rates. ``\textit{Not using batch API for data processing}'' was detected in 48\% of repositories, suggesting that nearly half of the projects process data inefficiently, potentially increasing latency and cloud costs. ``\textit{Improper handling of ML API limits}'' was observed in 36\% of repositories, indicating that many developers do not explicitly manage rate limits, which could lead to system failures or degraded performance when API quotas are exceeded.  
Less frequent misuses include ``\textit{Non-specification of early stopping criteria}'' (24\%, 199 occurrences), ``\textit{Not using training checkpoints}'' (17\%, 145 occurrences), and ``\textit{Misinterpreting output}'' (1.5\%). Although less prevalent, these misuses can still result in inefficiencies in model training, leading to longer training times and higher computational costs.

\begin{tcolorbox}[top=2pt, bottom=2pt, 
  left=2.2pt, right=2.2pt, boxsep=0pt, ]\small
  \textbf{Answer to RQ3:} There is a widespread and diverse presence of ML service misuse across the projects, with ``\textit{Ignoring monitoring for data drift}'' and ``\textit{Ignoring testing schema mismatch}'' being the most common, occurring in 98\% and 97\% of the projects, respectively.
\end{tcolorbox}

\vspace{1pt}
\section{Discussion}
\label{Sec:implications}
We describe the limitations of {\approach} and the implications of our results for researchers and practitioners.

\vspace{4pt}
\noindent\textbf{Implications for Researchers and Practitioners.} Our findings have important implications for researchers and practitioners.
For practitioners, {\approach} offers a concrete and effective solution for detecting and addressing ML service misuses early in the development lifecycle of ML service-based systems. Integrating {\approach} into CI/CD pipelines, for example, enables organizations to proactively identify and correct potential misuses before they propagate, directly improving software quality. Early detection can enhance maintainability, reduce costly post-deployment fixes, prevent performance degradation, and ensure consistent adherence to best practices for ML service integration.

For researchers, our work emphasizes the need for automated techniques to detect and mitigate ML service misuses, as few approaches target the specific misuse types we address in ML service-based systems. Unlike traditional ML systems, ML services have unique architectural and configuration constraints (e.g., service-specific APIs and hyperparameter tuning), giving rise to new misuse types not handled by existing tools. Our work fills this gap by identifying and detecting a novel class of misuses. The high prevalence of these misuses underscores the importance of raising awareness and improving best practices for ML services. Future research could leverage advanced ML techniques to detect more complex misuses and further refine detection algorithms. 

\vspace{3pt}
\noindent\textbf{Limitations of \approach. }Despite the effectiveness of {\approach} in detecting the seven ML services misuses, some limitations should be acknowledged. 
First, several detection algorithms in {\approach} rely on a manually curated list of cloud ML libraries, which we build by reviewing the official documentation of the three cloud providers supported by {\approach}. However, this manual curation process is time-consuming and prone to omissions, particularly for deprecated or newly introduced libraries. This also requires frequent updates for new ML service releases. To address these limitations, automated methods for tracking ML services changes as well as web scrapping of cloud providers documentation updates could be applied to maintain an up-to-date list and maintain the high detection accuracy of our approach. 

Second, {\approach} relies on static analysis and rule-based detection, which may not fully capture the dynamic behavior and runtime issues inherent in ML service-based systems. This limitation is also noted in other (anti)pattern detection approaches~\cite{tighilt2023maintenance,wan2021machine,cardozo2023prevalence,li2005pr}. Certain misuses, such as ``\textit{Ignoring monitoring for data drift}'' and ``\textit{Improper handling of ML API limits}'', could benefit from continuous monitoring and real-time analysis to improve detection, capabilities that our current implementation does not support. Despite this, {\approach} performs strongly using static analysis alone, achieving 96.7\% precision and 97\% recall, demonstrating its effectiveness in detecting ML service misuses. Integrating a hybrid detection approach that combines static and dynamic analysis represents a promising direction for future work. 

Lastly, {\approach} currently supports detecting seven ML service misuses across three major cloud providers. While effective, extending support to additional providers, services, and misuse types would increase its applicability as cloud ML offerings continue to grow. The metamodel and modular detection logic make {\approach} inherently extensible, allowing new providers, services, and misuses to be integrated with minimal effort. This design enables adaptation to evolving ML service ecosystems, improving detection coverage and maintaining effectiveness in real-world settings, while also supporting cross-provider comparisons and deeper understanding of common misuse patterns.

\section{Threats To Validity}
\label{Sec:threats}

\noindent\textbf{Internal validity.} One potential internal threat to the validity of our study is the reliance on a manually constructed ground truth for evaluating {\approach}. To mitigate this threat, multiple evaluators were involved to independently identify instances of misuses and minimize individual biases. Furthermore, we measured inter-rater agreement and observed a high level of consistency among evaluators which reinforces the reliability of our ground truth.

\vspace{2pt}
\noindent\textbf{Construct Validity.} To develop our approach, we leverage the \textit{ast} Python module, as to the best of our knowledge, there is currently no established method in the literature for parsing Python repositories specifically tailored to our needs. 
However, the ast module fails to parse files when there is a syntax error in the source code. To mitigate skipping important files, we implemented a preprocessing step that skips lines of code containing syntax errors. This step plays a crucial role in ensuring the robustness of the code analysis by filtering out lines that could introduce errors during parsing. 

\vspace{2pt}
\noindent\textbf{External validity.} {\approach} is currently designed for Python-based software systems, which may limit its applicability to other programming languages widely used in industry. However, Python is the language most commonly used for ML development, and we specifically targeted the three most prominent cloud providers in the ML cloud ecosystem. The reliance on a metamodel and dedicated detection algorithms makes {\approach} adaptable to other languages and extensible to additional misuses. While we demonstrate the effectiveness of {\approach} within this scope, broader validation is needed to improve the generalizability of our results. While static analysis cannot capture misuses introduced via Infrastructure-as-Code or runtime configuration, none of the 107 projects in our dataset relied on such mechanisms, limiting the impact of this limitation. Despite this, our static rules achieved high accuracy across all seven categories, demonstrating their effectiveness. A hybrid static-dynamic approach remains a promising direction for addressing runtime variability and edge cases. Future work will focus on extending support to other programming languages, expanding the range of detectable misuses, and supporting additional cloud providers to further enhance the generalizability of our results.

\section{Conclusion} 
\label{Sec:conclusion}
In this paper, we proposed \approach, a fully automated approach for detecting ML service misuses using a reusable metamodel and rule-based detection algorithms that identify seven common misuse types through static analysis of ML service-based systems. We evaluated {\approach} on 107 ML service-based systems and compared our results to a state-of-the-art baseline. {\approach} achieved an average precision of 96.7\% and a recall of 97\%, outperforming the only existing baseline. Applying {\approach} to 817 additional systems showed widespread ML service misuses, especially in data drift monitoring and schema validation, highlighting the need for better adherence to best practices in ML service integration. In future work, we plan to extend {\approach} to more ML service misuses and incorporate hybrid static-dynamic analysis to capture runtime issues. We also aim to add automated refactoring to fix detected misuses, improving ML service integration, code quality, and system reliability.

\section*{Acknowledgment}
This work is funded by the Natural Sciences and Engineering Research Council of Canada (NSERC): RGPIN-2023-05440 and RGPIN-2025-05897.

\balance
\bibliographystyle{unsrt}
\bibliography{paper}
\end{document}